\documentclass{wscpaperproc}
\pdfoutput=1
\usepackage{latexsym}
\usepackage{graphicx}
\usepackage{mathptmx}
\usepackage{textcomp}

\usepackage{amsmath}
\usepackage{amsfonts}
\usepackage{amssymb}
\usepackage{amsbsy}
\usepackage{amsthm}
\usepackage[pdftex,colorlinks=true,urlcolor=blue,citecolor=black,anchorcolor=black,linkcolor=black]{hyperref}

\newtheoremstyle{wsc}% hnamei
{3pt}% hSpace abovei
{3pt}% hSpace belowi
{}% hBody fonti
{}% hIndent amounti1
{\bf}% hTheorem head fontbf
{}% hPunctuation after theorem headi
{.5em}% hSpace after theorem headi2
{}% hTheorem head spec (can be left empty, meaning `normal')i

\theoremstyle{wsc}

\begin{document}

%***************************************************************************
% AUTHOR: AUTHOR NAMES GO HERE
% FORMAT AUTHORS NAMES Like: Author1, Author2 and Author3 (last names)
%
%		You need to change the author listing below!
%               Please list ALL authors using last name only, separate by a comma except
%               for the last author, separate with "and"
%

% setting up general page style
\pagestyle{fancyplain}

% setting up page style of first page
\thispagestyle{plain}
\firstPageHead{}

% setting up running header (authors) of subsequent pages
\chead{\fancyplain{}{\itshape An, Kim, Pakdamanian, and Brown}}

% setting up seperation parameters
%\headsep=72pt
\rhead{}
\cfoot{}
\renewcommand{\headrulewidth}{0pt} % (renewcommand needed in fancyhdr to remove top decorative line)
%\headrulewidth=0pt  % ("setlength" needed in fancyheading to remove top decorative line)

%%%%%%%%%%%%%%%%%%%%%%%%%%%%%%%%%%%%%%%%%%%%%%%%%%%%%%%%%%%%%%%%%%%%%%%%%%%%%%
%                                                                            %
%     THESE COMMANDS ARE REQUIRED TO WORK WITH WSC.BST TO MAKE BIBLIO     %
%                                                                            %
%%%%%%%%%%%%%%%%%%%%%%%%%%%%%%%%%%%%%%%%%%%%%%%%%%%%%%%%%%%%%%%%%%%%%%%%%%%%%%
\makeatletter
\let\@internalcite\cite
\def\cite{\def\@citeseppen{-1000}%
    \def\@cite##1##2{(##1\if@tempswa , ##2\fi)}%
    \def\citeauthoryear##1##2##3{##1 ##3}\@internalcite}
\def\citeNP{\def\@citeseppen{-1000}%
    \def\@cite##1##2{##1\if@tempswa , ##2\fi}%
    \def\citeauthoryear##1##2##3{##1 ##3}\@internalcite}
\def\citeN{\def\@citeseppen{-1000}%
%  Pierre L'Ecuyer's fix for multiple cite bug
%  Added by Paul J Sanchez on 4 October 2001
%   \def\@cite##1##2{##1\if@tempswa , ##2)\else{)}\fi}%
%   \def\citeauthoryear##1##2##3{##1 (##3}\@citedata}
    \def\@cite##1##2{##1\if@tempswa, ##2)\else{}\fi}%
    \def\citeauthoryear##1##2##3{##1 (##3)}\@citedata}
\def\citeA{\def\@citeseppen{-1000}%
    \def\@cite##1##2{(##1\if@tempswa , ##2\fi)}%
    \def\citeauthoryear##1##2##3{##1}\@internalcite}
\def\citeANP{\def\@citeseppen{-1000}%
    \def\@cite##1##2{##1\if@tempswa , ##2\fi}%
    \def\citeauthoryear##1##2##3{##1}\@internalcite}
\def\shortcite{\def\@citeseppen{-1000}%
    \def\@cite##1##2{(##1\if@tempswa , ##2\fi)}%
    \def\citeauthoryear##1##2##3{##2 ##3}\@internalcite}
\def\shortciteNP{\def\@citeseppen{-1000}%
    \def\@cite##1##2{##1\if@tempswa , ##2\fi}%
    \def\citeauthoryear##1##2##3{##2 ##3}\@internalcite}
\def\shortciteN{\def\@citeseppen{-1000}%
%  Pierre L'Ecuyer's fix for multiple cite bug
%  Added by Paul J Sanchez on 2 September 2002
%  should have caught this last year...
%   \def\@cite##1##2{##1\if@tempswa , ##2)\else{)}\fi}%
%   \def\citeauthoryear##1##2##3{##2 (##3}\@citedata}
% Shane G. Henderson fix for extra right bracket at end of optional material June 8, 2005
%    \def\@cite##1##2{##1\if@tempswa, ##2)\else{}\fi}%
    \def\@cite##1##2{##1\if@tempswa, ##2\else{}\fi}%
    \def\citeauthoryear##1##2##3{##2 (##3)}\@citedata}
\def\shortciteA{\def\@citeseppen{-1000}%
    \def\@cite##1##2{(##1\if@tempswa , ##2\fi)}%
    \def\citeauthoryear##1##2##3{##2}\@internalcite}
\def\shortciteANP{\def\@citeseppen{-1000}%
    \def\@cite##1##2{##1\if@tempswa , ##2\fi}%
    \def\citeauthoryear##1##2##3{##2}\@internalcite}
\def\citeyear{\def\@citeseppen{-1000}%
    \def\@cite##1##2{(##1\if@tempswa , ##2\fi)}%
    \def\citeauthoryear##1##2##3{##3}\@citedata}
\def\citeyearNP{\def\@citeseppen{-1000}%
    \def\@cite##1##2{##1\if@tempswa , ##2\fi}%
    \def\citeauthoryear##1##2##3{##3}\@citedata}
%
% \@citedata and \@citedatax:
%
% Place commas in-between citations in the same \citeyear, \citeyearNP,
% \citeN, or \shortciteN command.
% Use something like \citeN{ref1,ref2,ref3} and \citeN{ref4} for a list.
%
\def\@citedata{%
    \@ifnextchar [{\@tempswatrue\@citedatax}%
                  {\@tempswafalse\@citedatax[]}%
}

\def\@citedatax[#1]#2{%
\if@filesw\immediate\write\@auxout{\string\citation{#2}}\fi%
  \def\@citea{}\@cite{\@for\@citeb:=#2\do%
    {\@citea\def\@citea{, }\@ifundefined% by Young
       {b@\@citeb}{{\bf ?}%
       \@warning{Citation `\@citeb' on page \thepage \space undefined}}%
{\csname b@\@citeb\endcsname}}}{#1}}%

% don't box citations, separate with ; and a space
% also, make the penalty between citations negative: a good place to break.
%
\def\@citex[#1]#2{%
\if@filesw\immediate\write\@auxout{\string\citation{#2}}\fi%
  \def\@citea{}\@cite{\@for\@citeb:=#2\do%
    {\@citea\def\@citea{; }\@ifundefined% by Young
       {b@\@citeb}{{\bf ?}%
       \@warning{Citation `\@citeb' on page \thepage \space undefined}}%
{\csname b@\@citeb\endcsname}}}{#1}}%

% (from apalike.sty)
% No labels in the bibliography.
%
\def\@biblabel#1{}
\makeatother

%\newlength{\bibhang}
%\setlength{\bibhang}{2em}

% Indent second and subsequent lines of bibliographic entries. Taken
% from openbib.sty: \newblock is set to {}.
% \renewcommand{\refname}{REFERENCES}

\newdimen\bibindent
\bibindent=0.0em
% SEC: was \def\thebibliography#1{\section*{\refname\@mkboth
% SEC: was   {\uppercase{\refname}}{\uppercase{\refname}}}\list
\def\thebibliography#1{\section*{\refname}\list
   {}{\settowidth\labelwidth{[#1]}
   \leftmargin\parindent
   \itemindent -\parindent
   \listparindent \itemindent
   \itemsep 0pt
   \parsep 0pt}
   \def\newblock{}
   \sloppy
   \sfcode`\.=1000\relax}

           % Set up BiBTeX macros

% needed to make the tex document look more like the word counterpart :-(
\setlength{\baselineskip}{12.7pt}

% AUTHOR: Enter the title, all letters in upper case
\title{EXPLORING GAZE BEHAVIOR TO ASSESS PERFORMANCE IN DIGITAL GAME-BASED LEARNING SYSTEMS}

% AUTHOR: Enter the authors of the article, see end of the example document for further examples

\author{Brian An\\ 
Inki Kim \\
Erfan Pakdamanian \\
Donald E. Brown \\ [12pt]
Department of Systems and \\
Information Engineering \\
University of Virginia\\
Charlottesville, VA 22904, USA\\
% Multiple authors are entered as follows.
% You may also need to adjust the titlevbox size in the preamble - search for titlevboxsize
}

\maketitle

\section*{ABSTRACT}
The recent growth of sophisticated digital gaming technologies has spawned an \$8.1B industry around using these games for pedagogical purposes. Though Digital Game-Based Learning Systems have been adopted by industries ranging from military to medical applications, these systems continue to rely on traditional measures of explicit interactions to gauge player performance which can be subject to guessing and other factors unrelated to actual performance. This study presents a novel implicit eye-tracking based metric for digital game-based learning environments. The proposed metric introduces a weighted eye-tracking measure of traditional in-game scoring to consider the mental schema of a player's decision making. In order to validate the efficacy of this metric, we conducted an experiment with 25 participants playing a game designed to evaluate Chinese cultural competency and communication. This experiment showed strong correlation between the novel eye-tracking performance metric and traditional measures of in-game performance.

\section{INTRODUCTION}
The recent and rapid growth of the computer gaming industry has also served as a catalyst for the growth of an \$8.1B industry around computer game-based pedagogical tools also known as digital game-based learning (DGBL) systems \shortcite{adkins2017}. DGBLs have gained popularity across a wide range of applications for a number of reasons but primarily due to their repeatability and cost-effective scalability. Despite becoming a critical component to the educational economy, there has yet to be discovered an effective and cross-domain method to measure performance within DGBLs.

Initial research in this area focused on the potential learning benefits of commercial off-the-shelf entertainment games, but recently many industries and researchers have transitioned to the development and evaluation of games specifically designed for learning \shortcite{subrahmanyam1994effect,connolly2012systematic}. A number of studies have aggregated and objectively evaluated games ability to meet specific learning objectives \shortcite{connolly2012systematic}. Common to many of these evaluation methods is the use of explicit user-response scores which may be susceptible to guessing, thus not accurately reflecting a player's knowledge and performance. Eye-gaze technology used in tandem with DGBLs provides a novel path to measure player performance.

%%In an effort to address these concerns, implicit measures of performance have become a growing area of research interest. Current research of these measures have focused on various physiological signals and their correlation to various user states \shortcite{bellotti2013assessment,bajaj2016neuroscience}. Specifically, studies have found correlations between neurological signals such as Electroencephalography(EEG) and functional magnetic resonance imagine(fMRI) to the metrics associated with player engagement and player flow\shortcite{janicke2011psychological,ninaus2014neurophysiological}. Despite the growth of research in this area, attributing particular neurological signals with spatial and temporal game interactions continues to be problematic due to the lack of attribution specificity in most neurological input signals.%%

Eye-tracking has been successfully used in activity recognition, affect detection, engagement, cognitive load assessment, and general learning \shortcite{bulling2011eye,courtemanche2011activity,jaques2014predicting,wang2014eye}. Modern eye-tracking technologies allow sophisticated passive process tracking with minimal cognitive load of more traditional process tracking methods such as think-aloud protocols \cite{lohse1996comparison}. Presupposing the validity of the Eye-Mind assumption \shortcite{just1980theory}, the high frequency spatial and temporal information available through eye-tracking present an avenue by which one can infer both conscious and subconscious cognitive processes that would be too ephemeral for active process tracking methods. The general Eye-Mind assumption asserts that the object of fixation is the focus of attention and cognitive processing. Widespread research has supported the Eye-Mind assumption, though a number of studies have also presented situations where it was shown to be questionable or insufficient  \shortcite{kiili2014evaluating} especially in situations when the visual scene is unrelated to the intended cognitive activity.

This study contributes to the growing body of research investigating the eye-tracking behaviors as they relate to performance within DGBLs. Specifically, we aim to determine whether a player's eye-tracking behavior correlates to their performance in order to substantiate a novel eye-tracking based method for evaluating player performance. During this study, we evaluate the eye-tracking data of players in a DGBL designed to evaluate participants in Chinese Cross Cultural Competence. 

\section{BACKGROUND}
Despite the advances in the operationalization of DGBLs, research scrutinizing the methods of measuring user performance has not demonstrated the same level of growth. A majority of DGBLs leverage post-game examinations of learning objectives in the form of pre/post testing \shortcite{bellotti2013assessment}. The actual implementation of the pre/post tests span a number of methods to include evaluating player's knowledge through questionnaires/surveys\shortcite{fishwick2010experimental,coffey2017efficacy}. A common derivation of this method is the Situational Judgment Tests (SJT) designed to use the learned knowledge in practical applications \shortcite{lane2013learning}. These methods have been deployed on a wide scale primarily for their ease of implementation, though they suffer from several notable shortcomings. Specifically, they fail to consider whether the pre-test influenced the post-test results as seen in the case of the Second China game experiment where the pre/post tests were the same test with the questions reordered \shortcite{coffey2017efficacy}. 

Psycho-physiological data as a means of assessing learning performance within DGBLs is a rapidly growing research area. Specifically, the use of eye-tracking has gained significant traction due to the advances in both eye-tracking technology as well as DGBL development tools. Companies such as Tobii, Pupil Labs, and SensoMotoric Instruments have significantly reduced the cost and increased the resolution of eye-tracking devices while also offering a variety of collection devices for diverse collection situations. Additionally, the proliferation of popular gaming stacks to include Unity3d and Unreal Engine have made it possible for smaller research efforts to develop high quality scalable game environments for minimal costs.

With this proliferation, various applications of eye-tracking in DGBLs have been explored to determine game effectiveness.  One specific research area targets the measurement of player engagement and flow through the interpretation of affect-based metrics \shortcite{renshaw2009towards,jaques2014predicting,d2012gaze}. A limitation, however, of affect-based measures such as pupil dilation is that they are generally one dimensional and exhibit a variable lag therefore providing limited insight into complex cognitive processes associated with learning \shortcite{wierda2012pupil}. 

In order to better investigate user performance, researchers have used fixation location and gaze paths to examine performance in DGBLs. Specifically, researchers have conducted experiments to identify differences in visual behavior based on skill level. Kickmeier-Rust et al. found significantly different gaze path and interaction strategies between high/low performance levels of teenagers playing a game designed to teach European geography \shortcite{kickmeier2011tracking}. In a study examining adolescent behaviors in attention enhancement therapy, Pascual et al. found that higher performance children exhibited lower fixation densities. Additionally, they were able to use ensemble machine learning methods on a combination of the gaze data and user interaction features to create a prediction model \shortcite{frutos2015assessing}.

More recently, researchers have begun to investigate the use of eye-tracking to understand the process and rate of learning through gaze metrics \shortcite{wu2014conceptual,jozsa2011find}. J{\'o}zsa et al. studied the possibility of assessing a learning curve through trends in the fixation duration and total visit duration between successive stimuli. The results, however, were mixed due to the variability of the stimuli \shortcite{jozsa2011find}.

The overall purpose of this research effort is to introduce and evaluate the efficacy of a novel eye-tracking measure of performance in social-interaction DGBL systems. In doing so, we contribute to the rapidly growing need for effective methods of measuring performance in DGBL systems.

\section{EYE-TRACKING MEASURE OF PERFORMANCE (ETMP)}
Social Interaction-based serious games often have players interact with computer-driven avatars through finite dialogue trees. When a conversation commences, players are expected to sequentially select one of several dialogue options in order to progress through a conversation. In some cases, the dialogue options are  either appropriate or inappropriate, but often, dialogue options have varying degrees of appropriateness scores which can have various impacts on the rapport that players can build with the non-player characters. In the latter case, there is an opportunity to leverage the range of appropriateness with visual attention in order to produce a weighted score of appropriateness as proposed below:
\[ Player_{Score} = \sum_{i=1}^{x}\sum_{j=1}^{y}s_{ij} f_{ij} \]
\\
where \(s_{ij}\) is the score of response option \(j\) in question \(i\), \(f_{ij}\) is the fixation proportion on response option \(j\) relative to the other dialogue options in question \(i\), \(y\) is the number of dialogue options in a conversation instance, and \(x\) is the total number of conversation instances in the conversation. By using fixation proportions instead of absolute fixation durations, the ETMP minimizes the impact of variable reading speeds of the participants.  Research has shown a gaze bias in which people fixate on options with higher ratings of subjective preference \shortcite{glaholt2011eye,shimojo2003gaze}. Assuming the validity of the previously mentioned Eye-Mind Theory \shortcite{just1980theory} and gaze bias findings, this proposed formulation would conceptually evaluate both the appropriateness of a response as well as consider the mental schema the player uses to determine his or her response. It is important to note that this methodology only captures the overt movements of visual attention as measured through eye-tracking and makes no assumptions of the covert movements of attention that would not be observable through eye movements \cite{duchowski2002breadth}. The gaze-based weighting of performance in this formulation results in higher scores for those players that spend more visual attention on the appropriate answers and lower scores for participants that dwell on less appropriate answers.

\section{RESEARCH QUESTIONS}
In order to investigate the feasibility and viability of a gaze-based measure of performance, this study examines how users distribute their visual attention across multiple dialogue options to progress conversations with avatars within a serious game designed to evaluate Cross Cultural Competence in Chinese Culture. The specific research questions and related hypotheses are as follows:
\\

RQ-1) How do users distribute their visual attention across dialogue options in social interaction serious games? Do they focus their attention on preferred options and less so on rejected options? 
\begin{itemize}
\item H$_{1a}$: The game participants will spend significantly more time fixating on selected answer choices than non-selected choices

\item H$_{1b}$: The higher-scoring participants will spend significantly more time fixating on culturally appropriate responses

\item H$_{1c}$: The advanced domain knowledge or metacognitive-ability participants will spend significantly more time fixating on higher-score responses
\end{itemize}

The intent of RQ-1 is to better understand the visual behavior of participants in a serious game designed to improve Cross Cultural Competence.  Specifically, we investigate whether the visual behavior in this game is comparable to that of other research findings in both visual behavior in multiple choice environments and other DGBL systems.
\\

RQ-2) Does the Eye-Tracking Measure of Performance correlate to other measures of performance? Does higher domain knowledge, gaze bias, or higher traditional game scoring correlate to higher scores in the ETMP?
\begin{itemize}
\item H$_{2a}$: The game participants with advanced domain knowledge or metacognitive skill will score significantly higher in the ETMP than those participants with lower domain knowledge or less skilled.

\item H$_{2b}$: The Participants who scored higher in the traditional game scoring will also score significantly higher in the ETMP.

\item H$_{2c}$: The participants who exhibited a gaze bias on their selected answer will score significantly higher in the ETMP.

\item H$_{2d}$: The participants who exhibited a gaze bias on appropriate answers will score significantly higher in the ETMP.
\end{itemize}

The intent of RQ-2 is to provide initial validation to the ETMP through correlation to other indicators of performance whether they be in-game measurements or independent measures of Cross-Cultural Competence (3C). 3C continues to be an active area of research without a clear consensus on a 3C theoretical construct \shortcite{chiu2013cross}. Though we discovered numerous 3C constructs and inventories in our review of the literature, we chose the Cultural Intelligence (CQ) inventory to assess an individual's 3C given promising results in recent independent validation efforts \shortcite{van2008development,matsumoto2013assessing}.
\\
\section{QUASI-EXPERIMENTAL DESIGN}
In an effort to operationalize and validate the ETMP, we conducted an experiment with a DGBL system. The following section describes the quasi-experimental design, setup and analysis approach of the experiment\cite{cook2002experimental}.

\subsection{Participants and Design}
25 participants (40\% male; M$_{age}$=20.67; SD$_{age}$=1.80) were recruited from the University of Virginia to participate in the experiment. All of the recruited  participants were students in the Chinese Language Department. Each student was administered the Cultural Intelligence (CQ) Inventory to gauge self-reported 3C. 6 of the 25 participated in the study exceeded our 25\% eye-tracking data loss threshold and thus had to be excluded from the analysis of the results resulting in a sample size of 19. 

\subsection{Materials and Measures}
The stimulus in this experiment is a custom developed first-person serious game placed in a Chinese university called the Chinese Cultural Dimension Training System (CCDTS) \shortcite{bayer2017simple}. In the game, the user is expected to collaborate with familiar and unfamiliar Chinese students and professors while working on a group project. As the story continues, users engage in various conversations concerning politics, cultural habits, and stereotypes concluding with the player coordinating meeting times with a fellow student and visiting office hours to clarify questions with the professor. The game begins with an initial in-game training segment which introduces the storyline, allows the user to become accustomed to the game controls, and provides specific instruction to Chinese culture. It is then followed by two evaluation scenes called \textit{Library} and \textit{Office Hours}. In order to adapt the CCDTS to collect eye-tracking data, the bounding areas of each dialogue response had to be enlarged and had preset as Areas-of-Interest within the Tobii Pro Studio eye-tracking software.

\subsubsection{Scene Descriptions}
In the \textit{Library} scene, the player must engage with a fellow student about the group project details and coordinate meeting times. In the \textit{Office Hours} scene, the player engages with a professor during Office Hours to get formal feedback and recommendations about their group project.

\subsubsection{In-Game Scoring}
During the development of the game, it was necessary to score and validate the cultural appropriateness of specific dialogue responses by means of interrater reliability. For this purpose Chinese cultural experts were recruited from the Chinese language department. The raters were asked to rate each response on a likert scale of 1 to 5 with 1 being the least appropriate and 5 being the most appropriate.  The ratings determined the explicit scores associated with each response. This score was used to measure the player's performance as this experiment was not designed for a pre-test to target learning gains. The CQ inventory administered pre-stimulus, however, allowed us to investigate any pre-game cultural competency. A more detailed discussion of the scoring methodology is described by Bayer et al.\cite{bayer2017simple}.

Given this rating system, a response with a score of either 4 or 5 was considered to be an appropriate score.  Additionally, if an interaction did not have a response scored as a 4 or 5, only the highest scored response in that interaction was considered as a 'appropriate score' in determining the total fixation durations.
%A screenshot of the game is shown in Figure %\ref{figure:fig2}.
%
%
%\begin{figure}[!t]
%	\centering	
%	\includegraphics[width=3.5in]{Screenshot}
%	\caption{In-game Library scene with Mandarin Dialogue}
%	\label{figure:fig2}
%\end{figure}

\subsection{Apparatus}
The stimulus was presented on a 33-in screen with a pixel resolution of 2560 by 1440. The CCDTS was played through an executable Unity file and eye-tracking data was recorded using Tobii Pro Studio V1.1. During testing, participants were seated approximately 65 cm away from the screen. The vertical placement of the screen was adjusted such that the center of the screen was at eye level of each participant.
Eye-tracking data was collected using the Tobii X3 120, a screen-mounted collection system.  The system was calibrated for each participant using the Tobii Pro Studio animated 8 point calibration system. Calibration accuracy was recorded to be within 0.6\textdegree  of visual angle for both axes of all participants.

\subsection{Procedure}
\begin{figure}[!t]
	\centering	
	\includegraphics[width=6in]{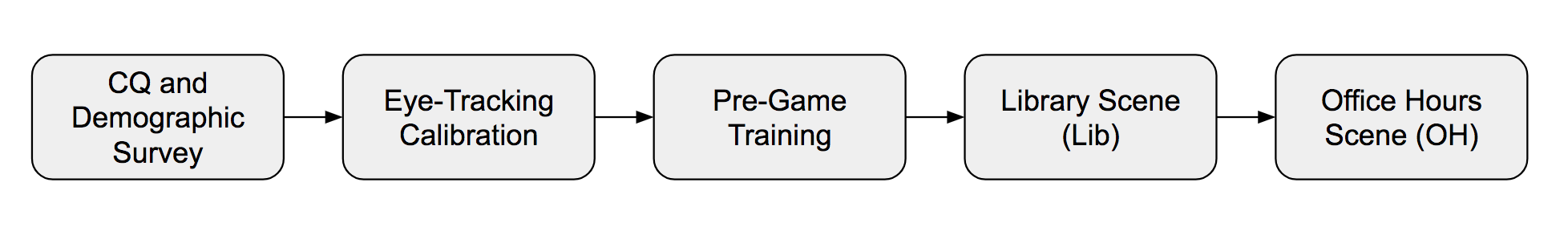}
	\caption{Experimental Design}
	\label{figure:fig1}
\end{figure}

Students were individually tested. Initially they were administered a demographic survey followed by the CQ Inventory. The CQ Inventory was used as a means to assess a players pre-game cross-cultural competence as a possible predictor of the ETMP. They were then calibrated to the eye-tracking apparatus within the tolerance of the eye-tracker and the CCDTS was initiated. The first scene in the game was a basic tutorial to teach participants how to navigate the interface and play the game. Once the tutorial was completed, the participant immediately entered the real game environment in which they interacted with non-player characters using the multiple choice style dialogue system common to modern video games. This test flow is depicted in Fig \ref{figure:fig1}. Following the completion of the game, a short after-action interview was conducted with the participant to attain any feedback about the game experience. The information collected from these interviews are not reported in this publication.

\subsection{Data Collection \& Analysis}
Prior to analyzing the data, the fixation durations on specific areas of interest (AOI) had to be processed from the collected gaze data. Since each interaction presented a series of responses for the participant to evaluate, the container around each of the responses was considered an AOI. All fixations within the container were aggregated to determine the total fixation duration in any particular response AOI.

Since each participant was exposed to different interactions based on their response choices, few participants saw the same sequence or set of interactions.  Additionally, each interaction had varying numbers of response choices meaning that the total fixation times for any particular interaction was expected to vary significantly since participants would presumably require more time to consider more responses. In order to account for the variability between subjects, \textit{percentage of fixation duration} was calculated for each interaction \shortcite{tsai2012visual}.

The primary methods used to test each of the hypotheses were paired t-test and multiple linear regression.

\section{RESULTS}
This section organizes the results by each hypothesis.

\subsection{H$_{1a}$: Participants will spend significantly more time fixating on selected answer choices than non-selected choices}

To examine this hypothesis, four Bonferroni-adjusted post-hoc paired t-tests were conducted on the percentage of fixation duration. The first paired t-test conducted compared the percentage of fixation duration on the selected options and the combined fixation duration percentage of non chosen options. The second paired t-test conducted compared the percentage of fixation duration on the selected option and the average percentage of non chosen options. This set of paired t-tests was individually conducted on the data of each scene.

The paired t-test for the Library scene comparing the percentage of fixation duration on selected options(\textit{M}=0.2246, \textit{SD}=0.06645)  and the combined duration percentage(\textit{M}=0.3117, \textit{SD}=0.1055) was significantly different(\textit{t}=-4.2552, \textit{df}=18, \textit{p}=0.00048). The paired t-test for the Libary scene comparing the percentage of fixation duration on selected options(\textit{M}=0.2426, \textit{SD}=0.06645)  and the average duration percentage(\textit{M}=0.1232, \textit{SD}=0.084) was significantly different(\textit{t}=7.4842, \textit{df}=18, \textit{p}=6.247e-07).

The paired t-test for the Office Hours scene comparing the percentage of fixation duration on selected options(\textit{M}=0.1985, \textit{SD}=0.054)  and the combined duration percentage(\textit{M}=0.4224, \textit{SD}=0.108) was significantly different(\textit{t}=-7.613, \textit{df}=18, \textit{p}=4.929e-07). The paired t-test for the Office Hours scene comparing the percentage of fixation duration on selected options(\textit{M}=0.1985, \textit{SD}=0.054)  and the average duration percentage(\textit{M}=0.12315, \textit{SD}=0.0298) was significantly different(\textit{t}=4.9635, \textit{df}=18, \textit{p}=0.0001005).

These results indicate that participants generally spent more time fixating on selected options than any other option.

\subsection{H$_{1b}$: Higher scoring participants will spend significantly more time fixating on culturally appropriate responses}

In general, participants spent more time fixating on the correct options(\textit{M}$_{Library}$=0.340, \textit{M}$_{Office Hours}$=0.346) rather than incorrect options(\textit{M}$_{Library}$=0.196, \textit{M}$_{Office Hours}$=0.275).

H$_{1b}$ was analyzed through linear regression employing the average Total Fixation on high scores as the dependent variable and the game score as the independent variable.  The average Total Fixation on high scores as defined previously was calculated by summing the fixations on high scores for a participant and dividing by the number of interactions the participant was exposed to throughout the game play labelled in the model as ``App Fix Lib" and ``App Fix OH" in Table \ref{tab1}. This analysis was performed separately on the Library scene and the Office Hours scene. 

The Library scene model did not show the Game Score as being a significant predictor for the average Total Fixation on high scores.  The Office Hours scene did show the Game Score as being a significant predictor for the average Total Fixation on high scores. The positive coefficient in this model indicates that higher game scores in the Office Hours scene were correlated to higher fixation duration percentages on high score responses.

% Scene 4 & 5 Regression Table
% Table created by stargazer v.5.2 by Marek Hlavac, Harvard University. E-mail: hlavac at fas.harvard.edu
% Date and time: Thu, Apr 12, 2018 - 03:04:17
\begin{table}[!htbp] \centering 
  \caption{Regression Analysis for H$_{1b}$} 
  \label{tab1}
\begin{tabular}{@{\extracolsep{5pt}}lcc} 
\\[-1.8ex]\hline 
\hline \\[-1.8ex] 
 & \multicolumn{2}{c}{\textit{Dependent variable:}} \\ 
\cline{2-3} 
\\[-1.8ex] & App\_Fix\_Lib & App\_Fix\_OH \\ 
\hline \\[-1.8ex] 
 Game\_Score\_Lib & 0.006 (0.007) &  \\ 
  Game\_Score\_OH &  & 0.008$^{**}$ (0.004) \\ 
  Constant & 0.202 (0.163) & 0.166  (0.081) \\ 
 \hline \\[-1.8ex] 
Observations & 19 & 19 \\ 
Adjusted R$^{2}$ & $-$0.015 & 0.186 \\ 
Residual Std. Error (df = 17) & 0.101 & 0.063 \\ 
F Statistic (df = 1; 17) & 0.738 &  5.118$^{**}$ \\ 
\hline 
\hline \\[-1.8ex] 
\textit{Note:}  & \multicolumn{2}{r}{$^{**}$p$<$0.05; $^{***}$p$<$0.01} \\ 
\end{tabular} 
\end{table} 

\subsection{H$_{1c}$: The participants with advanced domain knowledge or metacognitive ability will spend significantly more time fixating on higher-score responses}

For the purposes of this hypothesis, higher domain knowledge and metacognitive ability are characterized by a participant's CQ score.  Specifically, domain knowledge is captured by the Cognitive factor of CQ and metacognitive ability is captured by the Metacognitive factor of CQ. Based on a regression analysis where the dependent variable was the total Average fixation duration percentage on appropriate answers, both CQ factors were not shown to be significant as shown in Table \ref{tab2}.  Additionally, we analyzed the data to determine if the fixation duration percentage on inappropriate responses was correlated to the CQ factors, but this also did not produce statistically significant factors as shown in Table \ref{tab2}.

% Table created by stargazer v.5.2 by Marek Hlavac, Harvard University. E-mail: hlavac at fas.harvard.edu
% Date and time: Tue, Jun 26, 2018 - 15:25:34
\begin{table}[!htbp] \centering 
   \caption{Regression Results for H$_{1c}$ for Appropriate and Inappropriate Fixation Duration Percentage} 
   \label{tab2} 
\begin{tabular}{@{\extracolsep{5pt}}lcccc} 
\\[-1.8ex]\hline 
\hline \\[-1.8ex] 
 & \multicolumn{4}{c}{\textit{Dependent variable:}} \\ 
\cline{2-5} 
\\[-1.8ex] & App\_Fix\_Lib & App\_Fix\_OH & Inapp\_Fix\_Lib & Inapp\_Fix\_OH \\ 
\hline \\[-1.8ex] 
 CQ\_Metacognitive & 0.006 (0.035) & $-$0.010 (0.024) & $-$0.014 (0.028) & $-$0.022 (0.023) \\ 
  CQ\_Cognitive & $-$0.018 (0.032) & 0.026 (0.022) & 0.002 (0.026) & $-$0.001 (0.021) \\ 
  Constant & 0.400$^{*}$ (0.201) & 0.266$^{*}$ (0.136) & 0.262 (0.160) & 0.405$^{***}$ (0.134) \\ 
 \hline \\[-1.8ex] 
Observations & 19 & 19 & 19 & 19 \\ 
R$^{2}$ & 0.018 & 0.082 & 0.016 & 0.068 \\ 
Adjusted R$^{2}$ & $-$0.104 & $-$0.032 & $-$0.106 & $-$0.048 \\ 
Residual Std. Error (df = 16) & 0.106 & 0.071 & 0.084 & 0.070 \\ 
F Statistic (df = 2; 16) & 0.149 & 0.718 & 0.134 & 0.585 \\ 
\hline 
\hline \\[-1.8ex] 
\textit{Note:}  & \multicolumn{4}{r}{$^{*}$p$<$0.1; $^{**}$p$<$0.05; $^{***}$p$<$0.01} \\ 
\end{tabular} 
\end{table}

\subsection{H$_{2a}$: The participants with advanced domain knowledge or metacognitive skill will score significantly higher on the ETMP than those participants with lower domain knowledge.}

Again, domain knowledge and metacognitive skill were characterized by the specific CQ components described in H$_{1c}$. Not surprisingly, the Metacognitive and Cognitive component of CQ were not significant factors in both the Library and the Office Hours scene regression analysis as shown in Table \ref{tab4}.  When an Aikeke Information Criterion (AIC) based Stepwise regression was conducted, the CQ factors were eliminated \shortcite{akaike1974new}.

\subsection{H$_{2b}$: The participants who scored higher in the traditional game scoring will also score significantly higher on the ETMP.}

As hypothesized, participants' game scores in traditional scoring as described as ``Game Score Lib''(\textit{t}=2.884, \textit{df}=12, \textit{p}=0.01374) and ``Game Score OH''(\textit{t}=8.832, \textit{df}=12, \textit{p}=1.35e-06) in Table \ref{tab4} were positively correlated with the ETMP of both the Library and Office Hours scenes. The AIC-based Stepwise Regression also preserved these factors.

\subsection{H$_{2c}$: The participants who exhibited a gaze bias on their selected answers will score significantly higher on the ETMP.}

Participants who exhibited higher fixation duration percentages on their selected answers scored higher on the ETMP. The factor ``Avg answer fixation Lib''(\textit{t}=2.884, \textit{df}=12, \textit{p}=0.01374) as shown in Table \ref{tab4} represented the fixation duration percentage on selected answers for the Library scene. The factor ``Avg answer fixation OH''(\textit{t}=3.048, \textit{df}=12, \textit{p}=0.01034) as shown in Table \ref{tab4} represented the fixation duration percentage on selected answers for the Office Hours scene.

Additionally, the fixation duration percentage on the non-selected answers was also analyzed in the same model as described above. As seen in Table \ref{tab4} the variables ``Avg nonanswer fixation Lib''(\textit{t}=3.247, \textit{df}=12, \textit{p}=0.00699) and the ``Avg nonanswer fixation OH''(\textit{t}=4.291, \textit{df}=12, \textit{p}=0.00105) were also found to be significant.

\subsection{H$_{2d}$: The participants who exhibited a gaze bias on appropriate answers will score significantly higher on the ETMP.}
Participants who exhibited high fixation duration percentages on appropriate answers as defined in the results of H$_{1b}$ also resulted in higher ETMP scores as characterized by ``Appropriate Fix Lib'' (\textit{t}=1.846, \textit{df}=12, \textit{p}=0.08967) and ``Appropriate Fix OH''(\textit{t}=2.447, \textit{df}=12, \textit{p}=0.03076) in Table \ref{tab4}. Of these findings, ``Appropriate Fix Lib'' was found to be significant at the p$<$0.1 level which indicates a weak correlation and should be further investigated with a larger sample size experiment.

% Table created by stargazer v.5.2 by Marek Hlavac, Harvard University. E-mail: hlavac at fas.harvard.edu
% Date and time: Tue, Jun 26, 2018 - 15:35:11
\begin{table}[!htbp] \centering 
  \caption{Regression Analysis for Library and Office Hours ETMP} 
  \label{tab4} 
\begin{tabular}{@{\extracolsep{0pt}}lcccc} 
\\[-3ex]\hline 
\hline \\[-2.3ex] 
 & \multicolumn{4}{c}{\textit{Dependent variable:}} \\ 
\cline{2-5} 
\\[-2ex] & \multicolumn{2}{c}{ETMP Library} & \multicolumn{2}{c}{ETMP Office Hours} \\ 
\\[-3ex] & Original & Stepwise-AIC & Original & Stepwise-AIC\\  
\hline \\[-1.8ex] 
 CQ\_Metacognitive & $-$0.217 (0.345) &  & $-$0.192 (0.178) &  \\ 
  CQ\_Cognitive & 0.005 (0.327) &  & 0.179 (0.167) &  \\ 
  Avg\_answer\_fix\_Lib & 10.043$^{*}$ (5.525) & 10.406$^{*}$ (5.062) &  &  \\ 
  Avg\_nonanswer\_fix\_Lib & 14.869$^{***}$ (4.579) & 15.202$^{***}$ (4.273) &  &  \\ 
  App\_Fix\_Lib & 9.612$^{*}$ (5.207) & 9.208$^{*}$ (4.868) &  &  \\ 
  Game\_Score\_Lib & 0.252$^{**}$ (0.088) & 0.248$^{***}$ (0.082) &  &  \\ 
  Avg\_answer\_fix\_OH &  &  & 12.597$^{**}$ (4.133) & 11.712$^{**}$ (4.014) \\ 
  Avg\_nonanswer\_fix\_OH &  &  & 8.743$^{***}$ (2.037) & 8.699$^{***}$ (1.937) \\ 
  App\_Fix\_OH &  &  & 9.816$^{**}$ (4.011) & 10.817$^{**}$ (3.831) \\ 
  Game\_Score\_OH &  &  & 0.316$^{***}$ (0.036) & 0.315$^{***}$ (0.034) \\ 
  Constant & $-$4.158 (2.630) & $-$5.283$^{***}$ (1.674) & $-$5.280$^{***}$ (1.252) & $-$5.547$^{***}$ (0.823) \\ 
 \hline \\[-1.8ex] 
Observations & 19 & 19 & 19 & 19 \\ 
Adjusted R$^{2}$ & 0.904 & 0.914 & 0.962 & 0.963 \\ 
Residual Std. Error & 1.024 (df = 12) & 0.966 (df = 14) & 0.507 (df = 12) & 0.502 (df = 14) \\ 
F Statistic & 29.234$^{***}$  & 49.098$^{***}$  & 77.420$^{***}$ & 118.384$^{***}$ \\ 
Degrees of Freedom & (df = 6; 12) & (df = 4; 14) & (df = 6; 12) & (df = 4; 14)\\
\hline 
\hline \\[-1.8ex] 
\textit{Note:}  & \multicolumn{4}{r}{$^{*}$p$<$0.1; $^{**}$p$<$0.05; $^{***}$p$<$0.01} \\ 
\end{tabular} 
\end{table}

\section{DISCUSSION}
The findings from this study are organized by the research questions previously described.

The first research question investigates general visual behaviors with respect to social interaction serious games like the CCDTS. We found that participants did, in fact, fixate more on their selections rather than the non-selected choices.  This is consistent with various other studies that found longer fixation durations on selected options in multiple-choice based evaluation systems \shortcite{tsai2012visual,hegarty1992comprehension}. It is important to highlight, however, the large standard deviations of these results which may be an artifact of the varying numbers of responses for each set of dialogue options. This warrants future experiments with consistent quantities of options per interaction to investigate this hypothesis.

Additionally, we found a positive correlation between higher scoring participants and fixation duration on the more appropriate options in the Office Hours scene but not the Library scene. An explanation of this may be that the the Library scene was played prior to the Office Hours scene and, therefore, the user was visually calibrating his behavior resulting in more varied visual patterns. Assuming the Eye-Mind Theory, we interpret this result as being that higher scoring participants spent more time considering more appropriate answer options \cite{just1980theory} and less time considering less appropriate responses. That being the case, we postulate that concentrated visual and mental focus on appropriate responses rather than less appropriate responses more concretely indicates higher cultural competency. This then sets the foundation for the validation of the ETMP. 

Our analysis did not show significant correlation between our measures of domain knowledge or metacognition represented by Cultural Intelligence and fixation on appropriate answers. Though this measure has shown promise in various independent studies, the validation of this measure continues to be an active area of research and this finding motivates future experiments with other 3C constructs that have also shown potential\shortcite{chiu2013cross,matsumoto2013assessing}. 

With respect to the second research question exploring the validity of the ETMP as a measure of game performance, we found significant correlation between the ETMP and the other performance-related factors considered. Specifically, the linear regression models in Table \ref{tab4} revealed significant correlations to how much of their relative attention was focused on the selected and non-selected options listed as ``Avg answer fixation'' and ``Avg nonanswer fixation'', respectively. This specific finding possibly indicates that the ETMP takes into account the validity of the Eye-Mind Theory \cite{just1980theory}. Additionally, we also found a correlation between the ETMP and fixation on the appropriate answers in the game as indicated by the ``Appropriate Fix'' factor. This can be interpreted to mean that the ETMP also considers the accuracy of the mental schema that a player uses to determine his or her response though this interpretation is not entirely definitive. We also considered the possibility that rather this correlation may have occurred due to the low-score options being obviously socially unacceptable and easy to discard from consideration. This would then mean that the correct fixation was due in part to an understanding of general social norms rather than cultural competency. A targeted experiment with dialogue option sets that were more closely clustered in terms of appropriateness could resolve this interpretation. Finally, the ETMP was found to be strongly correlated to the player`s explicit game score which we interpret as being that the players' explicit performances were likely attributable to cultural proficiency. 

\section{LIMITATIONS}
The CCDTS dialogue structures were meant to facilitate a multitude of conversation tracks depending on the responses selected by the participants. Though this feature is ideal when considering realistic gameplay, it added complexity to the data structures and subsequently limited the sequential analysis that could be conducted since each player experienced different branches of the dialogue trees. For this reason, we were only able to effectively analyze the aggregated visual behavior metrics of each participant though we believe that a sequential analysis would provide insight into the specific metacognitive strategies employed by participants as well as performance improvement rates.

Also, despite instructions to maintain visual focus in the viewing area of the computer screen, several participants were observed consistently looking outside the limitations of the eye-tracker resulting in larger than expected track losses.  Future experiments may warrant head-mounted devices in order to maintain tracking throughout the experiment.

Additionally, given the construct of a linear storyline of the game it was impossible to introduce the two scenes in varying orders. As such it is plausible that there may exist some findings that are dependent on the scene such as the correlation to the appropriate responses as observed in Hypothesis H$_{1b}$. Future experiments could involve the development of games that were not dependent on this sequencing in order to isolate the results from scene order.

Finally, the stimuli used in this game and the experimental design were dependent upon the game construct. Specifically, the CCDTS was designed as a DGBL to teach and assess cross cultural competence to individuals with a baseline knowledge of the Mandarin language and Chinese culture and, as such, limited the available participant pool. Future experiments should target DGBL systems that allow for a broader range of participants to be able to generalize the findings beyond the specific application presented in this study.

\section{CONCLUSION}
In summary, this study introduces a novel eye-tracking measure of performance in social interaction serious games. We present a unique methodology by which players of DGBL systems can be assessed for not only their explicit in-game responses, but also consider the mental schema used to determine the selected response. This study also presents the results of an experiment whereby the ETMP was found to be correlated to gaze-behaviors indicative of specific cognitive processes and traditional measures of in-game performance. These findings contribute to the initial validation of the ETMP and motivate further research of the ETMP in other DGBL domains.

% Please don't exchange the bibliographystyle style
\bibliographystyle{wsc}
% AUTHOR: Include your bib file here
\bibliography{demobib}

\section*{AUTHOR BIOGRAPHIES}

\noindent {\bf BRIAN AN} is currently a Ph.D. candidate in the Department of Systems and Information Engineering, University of Virginia, Charlottesville, VA. His e-mail address is \email{baa4cb@virginia.edu}. \\

\noindent {\bf INKI KIM} is an Assistant Professor of Systems and Information Engineering at the University of Virginia. He holds a PhD in Industrial Engineering from the Pennsylvania State University. His research interests are in human behavior \& performance modeling and simulation-based training. In particular, design, analysis, and improvement of ``human(s)-in-the-loop`` systems that rely on streamlined human interaction with highly-automated machines are common theme of his research. His e-mail address is \email{ik3r@virginia.edu}.\\

\noindent {\bf ERFAN PAKDAMANIAN} is a Ph.D. student in the Department of  System and Information Engineering at the  University of Virginia. He received his M.S. in Industrial Engineering from Montana State University. His email address is \email{ep2ca@virginia.edu}.\\ 

\noindent {\bf DONALD E. BROWN} is Director of the Data Science Institute at the University of Virginia and William Stansfield Calcott Professor of Engineering and Applied Science. His research focuses on data fusion, predictive models, and simulation optimization. His email address is  \email{deb@virginia.edu}.\\

\end{document}